\documentclass[groupedaddress,aps,pra,superscriptaddress,showpacs,twocolumn,prl]{revtex4}%
\usepackage{epsfig,dsfont,amssymb,amsmath,amsthm,amsfonts,amsbsy,mathrsfs}
\usepackage{graphicx, color}
\usepackage{epstopdf}
\usepackage{mathdots}
\usepackage{float}
\begin{document}

\title{Basis-independent Coherence and its Applications}

\author{Zhi-Xiang Jin}
\affiliation{School of Computer Science and Technology, Dongguan University of Technology, Dongguan 523808, China}
\author{Yuan-Hong Tao}
\thanks{Corresponding author: taoyuanhong12@126.com}
\affiliation{School of Science, Zhejiang University of Science and Technology, 318 Liuhe Road, Hangzhou, Zhejiang 310023, China}
\author{Bing Yu}
\thanks{Corresponding author: mathyu590@163.com}
\affiliation{School of Mathematics and Systems Science, Guangdong Polytechnic Normal University, Guangzhou,510665,China}
\author{Shao-Ming Fei}
\thanks{Corresponding author: feishm@cnu.edu.cn}
\affiliation{School of Mathematical Sciences, Capital Normal University,  Beijing 100048,  China}
\affiliation{Max-Planck-Institute for Mathematics in the Sciences, Leipzig 04103, Germany}\vspace{7pt}

\begin{abstract}
In the quantitative theory of quantum coherence, the amount of coherence for given states can be meaningfully discussed only when referring to a preferred basis. One of the objections to this quantification is that the amount of coherence is an intrinsically basis-dependent quantity.
This limitation can, however, be lifted when considering a set of quantum states invariant under arbitrary unitary transformations. Thus, 
we analyze a basis-independent definition of quantum coherence, and the incoherent state is taken as the maximally mixed state. We describe the relationship between the basis-independent and the basis-dependent approaches and give several applications to show the advantages of the former method. 
The relations among basis-independent coherence, quantum entanglement, and quantum discord are discussed by using the relative entropy within a multipartite system. 

\end{abstract}
\maketitle
\section{Introduction}
Coherence, arising from the superposition principle of quantum states,  is one of the fundamental properties in quantum mechanics that distinguishes it from classical physics. 
One of the features of quantum coherence is that it can exist localized within an individual qubit, which contrasts it with other properties of quantum states such as entanglement. Besides, it can also be reviewed as correlations between the qubits \cite{cr}.
Thus, quantum coherence can be regarded as the prime ingredient required for various quantum technologies such as quantum state merging \cite{aes}, assisted subspace discrimination \cite{zfl} and quantum phase transitions \cite{gbf,agb, cx,szh} in many body systems.

A natural question is therefore how to characterize the coherence of quantum states. In Ref. \cite{pmb}, the authors present a rigorous method to quantify coherence by defining central concepts such as incoherent states and incoherent operations. 
Based on the framework introduced in Ref. \cite{pmb}, rapid developments have been made in a general resource theory of quantum coherence and its applications.  Similarly to the entanglement resource theory, investigations of general resource theory of quantum coherence mainly focused on new quantifiers, such as distance-based coherence measures \cite{pmb,cb,gc,rae,jzx1,rpp,ssd,reb,jzx,huml},  robustness of coherence \cite{nbc}, coherence cost \cite{wyd} and distillable coherence \cite{mx,csr}. While the application mainly focused on quantum simulation \cite{gan}, quantum metrology \cite{glm,dm}, and quantum cryptography \cite{eak}.

However, most of these quantifiers are due to the definition of what an “incoherent state” should be. That is
they are basis-dependent, which is a point of controversy and confusion, as the value of coherence in a physical system depends on the basis to be measured. Other quantum quantifiers to characterize a quantum state, for instance, quantum entanglement and quantum discord, generally are basis-independent. Different from the coherence measure quantified relative to a particular convention chosen by the observer, entanglement, and discord can be viewed as objective properties that the state possesses.  Several alternatives were suggested to overcome this point such as optimizing the quantum coherence over all possible local basis \cite{ycs,lsl1,yy,scp,yang,hml} 

In this paper, we analyze a basis-independent approach to measuring quantum coherence by modifying the set of incoherent states. We show that the only incoherent state with basis independent is the maximally mixed state \cite{ma}. That means any state that is not the maximally mixed state has coherence. Then we investigate the relation between the original coherence measure and basis-independent (BI) coherence measure in terms of relative entropy. It shows that given any state $\rho$, the BI coherence can decompose into original coherence under any basis, such as $\mathcal{B}$, and BI coherence of the measured state $\Phi^{\mathcal{B}}(\rho)$, where $\Phi^{\mathcal{B}}(\cdot)$ is a set of projection operators composed of $\mathcal{B}$.
Then we investigate the applications of BI coherence in wave-particle duality.
For states in multipartite systems, the relation among quantum entanglement, quantum discord, and BI coherence are presented.

\section{Basis-independent coherence measure}
In this section, we define a basis-independent measure of quantum coherence using relative entropy. In the original formulation of Ref. \cite{pmb}, quantum coherence is inherently a basis-dependent quantity, which measures the distance of a quantum state to the closest state in the set of incoherent states. This set of incoherent states is basis-dependent. For any state $\rho$, on one hand, one can always express $\rho$ in one basis such that it is incoherent (diagonal). On the other hand, a basic transformation on a diagonal state can convert it to a coherent state with off-diagonal terms.
Thus, it is significant to find out the diagonal states that remain incoherent under basis transformations.


We first define the free states, that is, those remaining diagonal states under basis transformations, featuring zero coherence. 
Assuming $\rho$ is a state in computational basis $\tilde{\mathcal{B}}:=\{|i\rangle\}_{i=1}^d$ and $T_j$ is the basis transition matrix from $\tilde{\mathcal{B}}$ to $\mathcal{B}_j:= \{|\alpha_i\rangle_j\}_{i=1}^d$, such that there exists a state $\rho$ for which all states $\rho_j=T^{-1}_j \rho T_j,~j=1,...,n$, in the bases $\{\mathcal{B}_j\}_{j=1}^n$ become diagonal, where $T^{-1}_j$ is the inverse matrix of $T_j$. Formally, the free states are given by 
\begin{eqnarray}\label{def}
F=\Big\{\rho|\rho_j=T^{-1}_j \rho T_j, \rho_j=\sum_{i=1}^d p_{i|j}|\alpha_i\rangle_j\langle\alpha_i|,~\forall j\Big\},
\end{eqnarray}
where $T_j$ is the transition matrix from basis $\tilde{\mathcal{B}}$ to $\mathcal{B}_j$ and $p_{\cdot|j}$ is a probability distribution for $j=1,...,n$. From the Eq.(\ref{def}), one can conclude if $\rho$ is a incoherent state, $\rho$ should commutate with each transition matrix $T_j,~j=1,...,n$, i.e., $\rho T_j= T_j\rho, ~\forall j$. As we all know, only scalar matrices can commutate with every matrix, so the maximally mixed state $I/d$ is the only basis-independent incoherent state, also see \cite{ma}.

Now let us consider a fixed density matrix $\rho$ in the $d$-dimensional Hilbert space $H$. Let $\Phi^{\mathcal{B}}(\rho)$ be a completely positive trace-preserving map on $\rho$ based on any basis $\mathcal{B}$. For a special 
observable $O$ with corresponding projectors $O_i=|i\rangle\langle i|$, $i=1,\cdots,d$ and $\tilde{\mathcal{B}}:=\{|i\rangle\}_{i=1}^d$ are the eigenvectors of the observable $O$, which form a complete orthogonal basis in Hilbert space $H$. $\Phi^{\tilde{\mathcal{B}}}(\rho)$ means the post-measurement state under projectors $\{O_i\}$, $i=1,\cdots,d$. 
We denote $C^\mathcal{B}(\rho)$ the original coherence based on any basis $\mathcal{B}$ \cite{pmb}, for example, $C^{\tilde{\mathcal{B}}}(\rho)$ is the original coherence in terms of the special basis $\{|i\rangle\}_{i=1}^d$, and $C(\rho)$ is the basis-independent (BI) coherence.

Let us consider the free set defined in Eq. (\ref{def}), taking a specific transition matrix such as $T_j= \mathbb{I}$ for simplicity. This reduces to the basis-dependent quantum coherence, which corresponds to having a fixed reference basis, namely, a computational one. The free set is given by $\tilde{F}=\big\{\rho|\rho=\sum_{i=1}^d p_i|i\rangle\langle i|\big\}$, and the relative entropy of coherence with respect to $\tilde{F}$ is defined as  
\begin{eqnarray}\label{rela}
C^{\tilde{\mathcal{B}}}(\rho)=\min_{\sigma\in \tilde{F}} S(\rho || \sigma)
\end{eqnarray}
with the quantum relative entropy $S(\rho || \sigma)=\mathrm{Tr}(\rho \log_2\rho)-\mathrm{Tr}(\rho \log_2\sigma)$.

Since the maximally mixed state $I/d$ is the only BI incoherent state, one can write the corresponding BI coherence measure explicitly as
\begin{eqnarray}\label{bn}
C(\rho)= S(\rho || I/d)=\log_2 d-S(\rho).
\end{eqnarray}
In contrast to the original relative entropy of coherence, a minimization over free states in equation $(\ref{rela})$ is not necessary due to the uniqueness of the free state in the resource theory of basis-independent coherence.
In Ref. \cite{la,hml}, it has been shown that under very generic conditions the normalized identity is the only classical state. As the identity $I/d$ is the only incoherence state that is diagonal in all bases, $C(\rho)$ can also be viewed as a quantifier of nonclassicality. In this sense, quantum coherence is more fundamental, which can be regarded as the basic unit to characterize quantumness.



In the following, we investigate the relation between BI coherence and original coherence. For any state $\rho$ in Hilbert space $H$, we have the following theorem, see proof in Appendix A.

{\bf Theorem 1}. Given a state $\rho$, for any basis $\mathcal{B}$, we have the following relation
\begin{eqnarray}\label{th1}
C(\rho)=C^\mathcal{B}(\rho)+C[\Phi^\mathcal{B}(\rho)],
\end{eqnarray}
where $\Phi^\mathcal{B}(\rho)$ is the post-measurement state under basis $\mathcal{B}$.

\begin{figure}
	\centering
	\includegraphics[width=11cm]{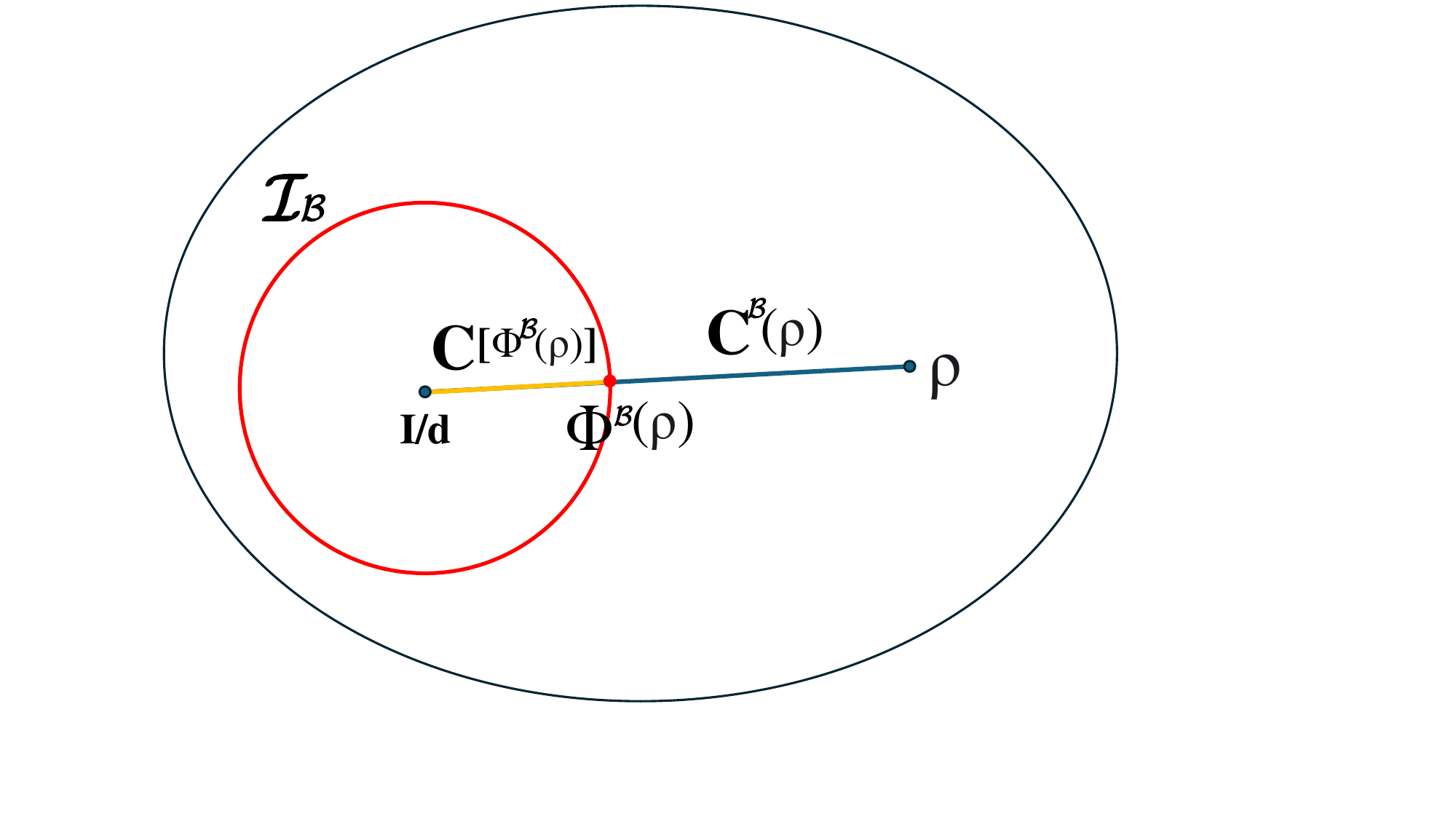}\\
	\caption{The red circle $\mathcal{I}_{\mathcal{B}}$ represents the set of all incoherent states under the bases $\mathcal{B}$. The orange line is the BI coherence of post-measure state $\Phi^{\mathcal{B}}(\rho)$ and the blue line is the original coherence under basis $\mathcal{B}$. The red dot is just the optimal solution of Eq.(\ref{rela})}\label{2}
\end{figure}

That is to say, given any state $\rho$, the BI coherence $C(\rho)$ can decomposite into the original relative entropy of coherence $C^\mathcal{B}(\rho)$ and BI coherence of the post-measurement state $C[\Phi^\mathcal{B}(\rho)]$. Moreover, the optimal solution of Eq.(\ref{rela}), i.e., $\Phi^\mathcal{B}(\rho)$, happens to be the intersection state of the line, between state $\rho$ and incoherent state $I/d$, with the set of incoherent states under bases $\mathcal{B}$, see Fig. 1. $C(\rho)$ is different from the intrinsic coherence defined in Ref. \cite{cr}, $C_I(\rho):=\min_{\sigma} S(\rho || \sigma)$ with $\sigma=\sum_{i}p_i\rho^i_1\otimes\rho^i_2\otimes\cdots\otimes\rho^i_n$ where $\rho^i_j$ is the incoherent state (not necessarily in one basis ) on the subsystem $j,~j=1,2,\cdots,n$, which in fact equal to the entanglement. In fact, they have the following order $C(\rho)\geq C^\mathcal{B}(\rho)\geq C_I(\rho)$.

$C[\Phi^\mathcal{B}(\rho)]$ in Eq. (\ref{th1}) plays a central role in bridging the quantification of original coherence under the reference basis $\mathcal{B}$ and the BI coherence. Under any specific basis $\mathcal{B}$, 
one gets $C[\Phi^\mathcal{B}(\rho)]=\log_2 d-S(\Phi^\mathcal{B}(\rho))$, which essentially characterizes the difference between the BI coherence and the original basis-dependent coherence measure $C^\mathcal{B}(\rho)$. We have $0\leq C[\Phi^\mathcal{B}(\rho)]\leq \log_2 d$ with the lower and upper bound obtained for $\rho$ being maximally mixed state and the one of the basis states, respectively.
 For a single qubit can be written as $\rho=\frac{I+\vec{r}\cdot \vec{\sigma}}{2}$, where $\vec{r}=(x,y,z)$ is a real three-dimensional vector such that $x^2+y^2+z^2\leq 1$, and $\vec{\sigma}=(\sigma_x, \sigma_y, \sigma_z)$ with $\sigma_i,~i=x,y,z,$ being pauli matrix. In particular, $\rho$ is pure if and only if $x^2+y^2+z^2=1$.
Under the computational basis $\tilde{\mathcal{B}}:=\{|i\rangle\}_{i=1}^d$ and basis  $\mathcal{B'}=\{\frac{|0\rangle+|1\rangle}{\sqrt{2}}, \frac{|0\rangle-|1\rangle}{\sqrt{2}}\}$, see Fig 2, the trade-off relations of $C^\mathcal{B}(\rho)$ and $C[\Phi^\mathcal{B}(\rho)]$ with respect to $\mathcal{B}=\{ \tilde{\mathcal{B}},\mathcal{B'} \}$ for $x=0.4$, see Appendix B for detailed derivations.
In order to see the relations between bases $\tilde{\mathcal{B}}$ and $\mathcal{B'}$ more clarity, we set $x=0.4$, $y=0.2$, see Fig 3. One can see that the sum of original coherence $C^\mathcal{B}(\rho)$ and BI coherence $C[\Phi^\mathcal{B}(\rho)]$ under different bases equals BI coherence $C(\rho)$.

In Eq.(\ref{th1}), we see that the decomposition of $C(\rho)$ depends on the basis as for $C^\mathcal{B}(\rho)$. In the following, we investigate the decomposition of $C(\rho)$ under any two sets of bases $\mathcal{B}_1$ and $\mathcal{B}_2$.

{\bf Corollary 2}. 
Given a state $\rho$, for any two bases $\mathcal{B}_1$ and $\mathcal{B}_2$, we have the following relation
\begin{eqnarray}\label{co2}
C(\rho)=C^{\mathcal{B}_1}(\rho)+C^{\mathcal{B}_2}[\Phi^{\mathcal{B}_1}(\rho)]+C[\Phi^{\mathcal{B}_2\mathcal{B}_1}(\rho)].
\end{eqnarray}
{\it Proof.} From theorem 1, one has 
\begin{eqnarray}\label{pfco21}
C(\rho)=C^{\mathcal{B}_1}(\rho)+C[\Phi^{\mathcal{B}_1}(\rho)]
\end{eqnarray}
under basis $\mathcal{B}_1$. Similarly, one can obtain 
\begin{eqnarray}\label{pfco22}
C[\Phi^{\mathcal{B}_1}(\rho)]=C^{\mathcal{B}_2}[\Phi^{\mathcal{B}_1}(\rho)]+C[\Phi^{\mathcal{B}_2\mathcal{B}_1}(\rho)]. 
\end{eqnarray}
Substituting Eq. (\ref{pfco22}) into Eq. (\ref{pfco21}), one obtains the conclusion.

Suppose $\mathcal{B}_1:=\{|\alpha_i\rangle\}$ and $\mathcal{B}_2:=\{|\beta_j\rangle\}$ are two sets of standard orthogonal bases. For any state $\rho$, one has $\rho=\sum_k \lambda_k |\psi_k\rangle\langle\psi_k|$, where $ \lambda_k$ and $|\psi_k\rangle$ are the eigenvalues and eigenvectors of state $\rho$, respectively. Then

\begin{eqnarray}\label{}
\rho&=&\sum_k \lambda_k |\psi_k\rangle\langle\psi_k|\nonumber\\
&=&\sum_k \lambda_k\sum_i c_{ki}|\alpha_i\rangle\sum_j c_{kj}\langle\alpha_j|\nonumber\\
&=&\sum_{i,j,k}\lambda_kc_{ki}c_{kj}|\alpha_i\rangle\langle\alpha_j|,
\end{eqnarray}
where $ |\psi_k\rangle=\sum_i c_{ki}|\alpha_i\rangle$ with $\sum_ic^2_{ki}=1$.

Then the postmeasured state under measurement $\{|\alpha_i\rangle\langle\alpha_i|\}$ is
 \begin{eqnarray}\label{}
\Phi^{\mathcal{B}_1}(\rho)&=&\sum_l |\alpha_l\rangle\langle\alpha_l|\rho|\alpha_l\rangle\langle\alpha_l|\nonumber\\
&=&
\sum_l |\alpha_l\rangle\langle\alpha_l|\sum_{i,j,k}\lambda_kc_{ki}c_{kj}|\alpha_i\rangle\langle\alpha_j|\alpha_l\rangle\langle\alpha_l|\nonumber\\
&=&\sum_{i,k}\lambda_kc^2_{ki}|\alpha_i\rangle\langle\alpha_i|.
 \end{eqnarray}
If $\Phi^{\mathcal{B}_1}(\rho)=I/d$, then $\sum_k \lambda_kc^2_{ki}=1/d$ must be satisfied, i.e., $\lambda_k=1/d,~ \forall k$. 
If $\Phi^{\mathcal{B}_1}(\rho)\neq I/d$, the measured state $\Phi^{\mathcal{B}_1}(\rho)$ under the measurement $\{|\beta_j\rangle\langle\beta_j|\}$ is
 \begin{eqnarray}\label{}
\Phi^{\mathcal{B}_2\mathcal{B}_1}(\rho)&=&\sum_j |\beta_j\rangle\langle\beta_j| \sum_{i,k}\lambda_kc^2_{ki}|\alpha_i\rangle\langle\alpha_i| |\beta_j\rangle\langle\beta_j|\nonumber\\
&=&
\sum_{i,j,k} \lambda_kc^2_{ki}|\langle\beta_j|\alpha_i\rangle|^2|\beta_j\rangle\langle\beta_j|. 
\end{eqnarray}
If $\mathcal{B}_1$ and $\mathcal{B}_2$ are MUB, then $|\langle\beta_j|\alpha_i\rangle|^2=\frac{1}{d},~\forall i,j$. Thus  
 \begin{eqnarray}\label{}
\Phi^{\mathcal{B}_2\mathcal{B}_1}(\rho)&=&\sum_{i,j,k} \lambda_kc^2_{ki}|\langle\beta_j|\alpha_i\rangle|^2|\beta_j\rangle\langle\beta_j|\nonumber\\
&=&\frac{1}{d}\sum_{i,j,k} \lambda_kc^2_{ki}|\beta_j\rangle\langle\beta_j|\nonumber\\
&=&\frac{1}{d}\sum_j|\beta_j\rangle\langle\beta_j|.
\end{eqnarray}

That is to say, $C[\Phi^{\mathcal{B}_2\mathcal{B}_1}(\rho)]=0$ under the conditions of $\mathcal{B}_1$ and $\mathcal{B}_2$ being  mutually unbiased bases (MUBs), then Eq. (\ref{co2}) reduced to $C(\rho)=C^{\mathcal{B}_1}(\rho)+C^{\mathcal{B}_2}[\Phi^{\mathcal{B}_1}(\rho)]$.
If $\mathcal{B}_1$ and $\mathcal{B}_2$ are the same, then we have $C(\rho)=C^{\mathcal{B}_1}(\rho)+C[\Phi^{\mathcal{B}_1}(\rho)]$ since $C^{\mathcal{B}_1}[\Phi^{\mathcal{B}_1}(\rho)]=0$, i.e., Eq. (\ref{co2}) reduced to theorem 1. Otherwise, one gets $C(\rho)>C^{\mathcal{B}_1}(\rho)+C^{\mathcal{B}_2}[\Phi^{\mathcal{B}_1}(\rho)]$ since $C[\Phi^{\mathcal{B}_2\mathcal{B}_1}(\rho)]>0$.
Generally, for any $n$ sets of bases $\{\mathcal{B}_1,\mathcal{B}_2,\cdots, \mathcal{B}_n\}$, we have the following result.

\begin{figure}
	\centering
	\includegraphics[width=7cm]{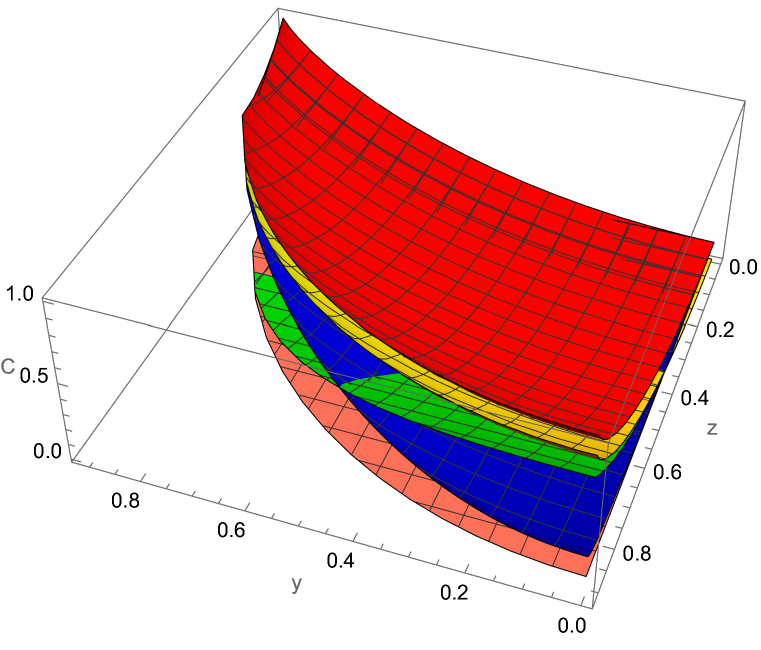}\\
	\caption{The decomposition of BI coherence $C$ under basis $\tilde{\mathcal{B}}$ and $\mathcal{B'}$ with $x=0.4$. The red curved surface is the BI coherence. The blue curved surface represent $C^{\tilde{\mathcal{B}}}(\rho)$ under basis $\tilde{\mathcal{B}}$ and green curved is the BI coherence $C[\Phi^{\tilde{\mathcal{B}}}(\rho)]$. The yellow curved surface represent $C^\mathcal{B'}(\rho)$ under basis $\mathcal{B'}$ and pink curved surface is the BI coherence $C[\Phi^\mathcal{B'}(\rho)]$.}\label{2}
\end{figure}

\begin{figure}
	\centering
	\includegraphics[width=7cm]{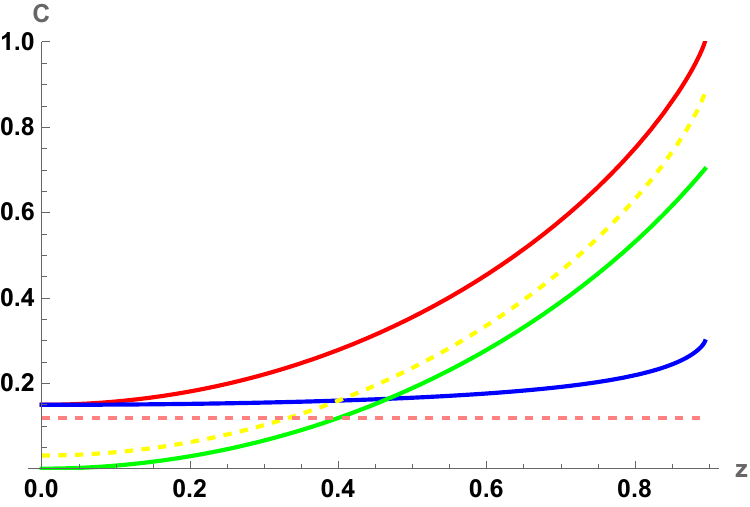}\\
	\caption{The decomposition of BI coherence $C$ under basis $\tilde{\mathcal{B}}$ and $\mathcal{B'}$ with $x=0.4$, $y=0.2$. One can see that the pink dashed line is an straight line with $C[\Phi^\mathcal{B'}(\rho)]=0.1187$. The yellow dashed line is the $C^\mathcal{B'}(\rho)$.The blue solid line is $C^{\tilde{\mathcal{B}}}(\rho)$ and green solid line is $C[\Phi^{\tilde{\mathcal{B}}}(\rho)]$.}\label{2}
\end{figure}

{\bf Corollary 3}. 
Given a state $\rho$, for $n$ sets of bases $\{\mathcal{B}_1,\mathcal{B}_2,\cdots, \mathcal{B}_n\}$, we have the following relation
\begin{eqnarray*}\label{co3}
	C(\rho)&=&C^{\mathcal{B}_1}(\rho)+C^{\mathcal{B}_2}[\Phi^{\mathcal{B}_1}(\rho)]+\cdots\nonumber\\
	&+&C^{\mathcal{B}_n}[\Phi^{\mathcal{B}_{n-1}\mathcal{B}_{n-2}\cdots \mathcal{B}_1}(\rho)]+C[\Phi^{\mathcal{B}_{n}\mathcal{B}_{n-1}\cdots \mathcal{B}_1}(\rho)].
\end{eqnarray*}

In fact, the worst case is that $\Phi^{\mathcal{B}_n\mathcal{B}_{n-1}\cdots\mathcal{B}_1}(\rho)\ne I/d$.
\begin{eqnarray}\label{}
&&\Phi^{\mathcal{B}_n\mathcal{B}_{n-1}\cdots\mathcal{B}_1}(\rho)\nonumber\\
&&=\sum_{k,i_1,\cdots,i_n} \lambda_kc^2_{ki}|\langle\alpha^2_{i_2}|\alpha^1_{i_1}\rangle|^2\cdots \nonumber\\ &&|\langle\alpha^{n-1}_{i_{n-1}}|\alpha^{n-2}_{i_{n-2}}\rangle|^2|\langle\alpha^{n}_{i_{n}}|\alpha^{n-1}_{i_{n-1}}\rangle|^2|\alpha^{n}_{i_{n}}\rangle\langle\alpha^{n}_{i_{n}}|,
\end{eqnarray}
where $\Phi^{\mathcal{B}_j}(\rho)$ means the $j$-th measurement  $\{|\alpha^j_{i_j}\rangle\langle \alpha^j_{i_j}|\}$ acts on state $\rho$, i.e., $\sum_{k,i_1,\cdots,i_{n-1}} \lambda_kc^2_{ki}|\langle\alpha^2_{i_2}|\alpha^1_{i_1}\rangle|^2\cdots|\langle\alpha^{n}_{i_{n}}|\alpha^{n-1}_{i_{n-1}}\rangle|^2\ne \frac{1}{d}$. 
$C[\Phi^{\mathcal{B}_n\mathcal{B}_{n-1}\mathcal{B}_1}(\rho)]$  is a monotonically non-increasing function, since $C^{\mathcal{B}_i}[\Phi^{\mathcal{B}_{i-1}\mathcal{B}_1}(\rho)]\geq 0$, $i=2,\cdots,n$. If $\{\mathcal{B}_1,\mathcal{B}_2,\cdots,\mathcal{B}_n\}$ runs over all bases, then $\Phi^{\mathcal{B}_n\mathcal{B}_{n-1}\cdots\mathcal{B}_1}(\rho)=I/d$. This does not mean $\Phi^{\mathcal{B}_n\mathcal{B}_{n-1}\cdots\mathcal{B}_1}(\rho)=I/d,~n\to \infty$. For example, one can choose one set of incoherent states under any basis, such as $\mathcal{B}_1$, and then regard I/d as a rotation axis; one can obtain infinite sets of incoherent states as the axis rotates. The final state $\Phi^{\mathcal{B}_n\mathcal{B}_{n-1}\cdots\mathcal{B}_1}(\rho)$ (red point a in Fig. 4) is just the intersection state of the line connecting the initial state $\rho$ and BI incoherent state I/d with the set of incoherent states under basis $\mathcal{B}_n$, see Fig.4. From the above analysis, for any state $\rho$, one can transform $\rho$ to incoherent state $I/d$ by using at least two sets of measurement, and the worst case is that even infinite measurements cannot be converted into incoherent state $I/d$.
\begin{figure}
	\centering
	\includegraphics[width=9.5cm]{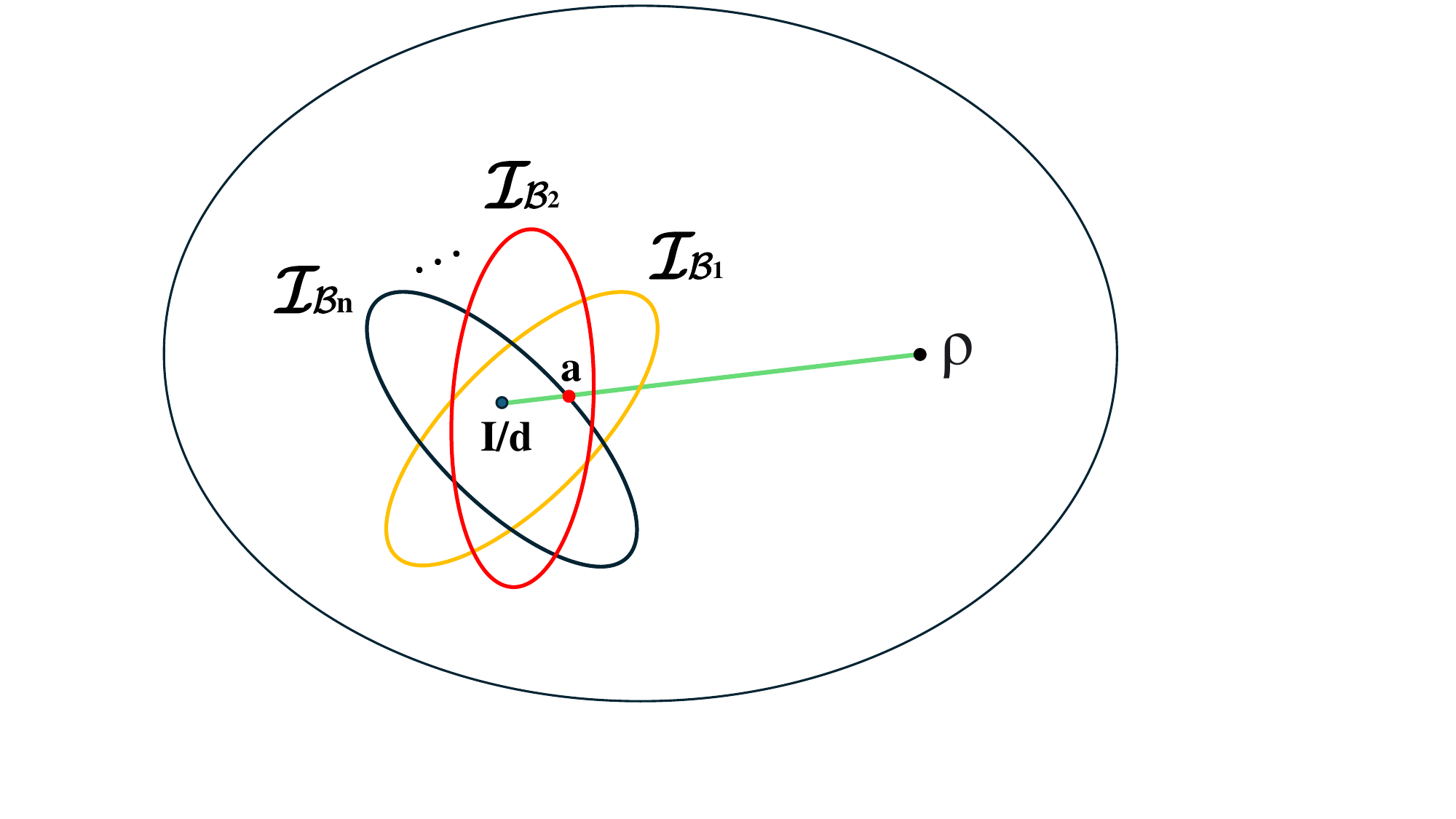}\\
	\caption{The yellow, red, blue circle $\mathcal{I}_{\mathcal{B}_1}$, $\mathcal{I}_{\mathcal{B}_2}$, $\cdots$, $\mathcal{I}_{\mathcal{B}_n}$ represents the set of incoherent states under the bases $\mathcal{B}_1$, $\mathcal{B}_2$, $\cdots$, $\mathcal{B}_n$. The green line is the BI coherence of original state $\rho$. The red point a is the final state $\Phi^{\mathcal{B}_n\mathcal{B}_{n-1}\cdots\mathcal{B}_1}(\rho)$.}\label{2}
\end{figure}

This raises us to consider another interesting question: the maximally coherent states such that $C(\rho)$ is maximal. From Eq. (\ref{bn}), $C(\rho)\leq \log_2d$ and the equation satisfied for all pure states, which are maximally coherent states since $S(|\psi\rangle)=0$ with $|\psi\rangle$ a pure state. 
From Eq. (\ref{co2}), one has $C^\mathcal{B}(\rho)\leq C(\rho)$. That is to say, for any state $\rho$, the original coherent measure is no more than the BI coherent measure. In the following, we investigate the state $\rho$ such that $C^\mathcal{B}(\rho)=C(\rho)$. For pure states, it is easy to know that if the state $\rho$ is the maximum superposition state under the basis $\mathcal{B}$, then $C^\mathcal{B}(\rho)=C(\rho)$.
For mixed states,
suppose $\rho=\sum_i p_i|i\rangle\langle i|$ under the basis $\mathcal{B}_1$, if $p_i=\frac{1}{d}, \forall i$, then $C(\rho)=0$, the reverse also established. Otherwise, we take another basis $\mathcal{B}_2:=\{|i'\rangle \}$ such that $\mathcal{B}_1$ and $\mathcal{B}_2$ are MUBs. i.e., $|\langle i|j'\rangle|=\frac{1}{\sqrt{d}}$. Then $\rho'=\sum_jp_j|j'\rangle\langle j'|$ is a maximally coherent mixed state for original coherent measure based on $\mathcal{B}_1$, since $C^{\mathcal{B}_1}(\rho')=C(\rho)$.
Thus, given a state $\rho$, $C^\mathcal{B}(\rho)$ and $C[\Phi^\mathcal{B}(\rho)]$ have a trade-off relation, $C^\mathcal{B}(\rho)$ decreasing as $C[\Phi^\mathcal{B}(\rho)]$ increasing, and $C^\mathcal{B}(\rho)=0$ using the eigenvectors of $\rho$ as the basis $\mathcal{B}$, while reaches maximum $\log_2 d-S(\rho)$ for any bases that MUB with $\mathcal{B}$.

\section{Applications of the BI coherence in wave-particle duality}
Since quanton displays particle or wave nature is dependent on the measurement apparatus, Eq. (\ref{th1}) can be used to quantify wave and particle properties. The off-diagonal elements of state $\rho$ determine wave interference, while the diagonal elements of state $\rho$ determine the distinguishability of path information. 

Suppose $f$ characterize a valid particle property, the function $f$ should satisfy the following criteria \cite{durr}: 1).~$f$ should reach its global maximum iff $\rho_{ii}=1$ for one
$i$, i.e., the path is certain; 2).~$f$ should reach its global minimum iff $\rho_{ii}=1/d$ for all
$i$, i.e., the path is completely uncertain; 3).~$f$ is invariant under permutations of the $d$ path labels; 4).~$f$ is convex, namely, for any two density matrices $\rho_1$ and $\rho_2$, one has for $\rho=(1-\lambda)\rho_1+\lambda\rho_2$ ($0\leq\lambda\leq1$),
$f(\rho^{\mathrm{diag}})\leq(1-\lambda) f(\rho_{1}^{\mathrm{diag}})+\lambda f(\rho_{2}^{\mathrm{diag}})$, where $\rho^{\mathrm{diag}}$ is the diagonal element of the density matrix.
One can take relative entropy as the function $f$, then define the measure of particle as  \begin{eqnarray*}
	P^{\tilde{\mathcal{B}}}(\rho):=C(\rho^{\mathrm{diag}})=\log_2 d-S(\rho^{\mathrm{diag}}).
\end{eqnarray*}
 It is easy to conclude that $C(\rho^{\mathrm{diag}})$ satisfies the above first three items. As for the last item, due to von Neumann entropy being a concave function, one can obtain the desired result.

\if
If the initial state is $|o_i\rangle$, one knows the outcome with certainty even before a measurement of $O$ is performed. Thus the quanton is in a definite path.
Suppose that the initial state is $\rho$, and Alice measures the observable $O$, then delivers the quanton to Bob without telling him the outcome. Aware of the observable measured, Bob is certain that in each run he receives a state $\rho_i=O_i$ under a definite path with probability $p_i=\mathrm{Tr}(O_i\rho O_i)$. However, without accessing any information about the outcomes, Bob predicted that the state he received is a statistical mixture of definite paths, $\sum_i p_i\rho_i$. By the theorem $S(\sum_jp_j\rho_j)=H(p_j)+\sum_j p_j S(\rho_j)$, where $S$ and $H$ stand for the von Neumann and Shannon entropies, respectively, we see that $S(\sum_ip_iO_i)=H(p_i)=-\sum_ip_i \ln p_i$. That is, Bob's lack of knowledge is just the classical probability distribution $p_i$ secretly prepared by Alice.

These aspects naturally fit our classical intuition by which a quanton always is on a definite path, even when one does not know any information about this path. Now, considering a quanton is in a definite path state, its property will not be disturbed by a projective measurement on that path.
This motivates us to define particle property as
\begin{eqnarray*}
\Phi^{\mathcal{B}_O}(\rho)=\rho,
\end{eqnarray*}
where $\Phi^{\mathcal{B}_O}(\rho):=\sum_i^d\langle o_i|\rho|o_i\rangle O_i$, here $\mathcal{B}_O:=\{|o_i\rangle\langle o_i|,i=1,2,...,d\}$. If a state $\rho$ is particle property with respect to $\mathcal{B}_O$, then it follows that $f(\rho)=f(\Phi^{\mathcal{B}_O}(\rho))=f(\rho_{diag})$ for a generic function $f$ with $\rho_{diag}$ a matrix composed of diagonal elements of $\rho$.

\fi

Correspondingly, the wave aspect is characterized by the off-diagonal elements of $\rho$.
As a well-defined measure of the wave aspect, the wave property $g(\rho)$ should satisfy the following conditions: 1). $g$ should reach its global minimum iff $\rho=\rho^{\mathrm{diag}}$; 2). $g$ should reach its global maximum iff $\rho=\sum_{j,k}|j\rangle \langle k|{\rm e}^{\mathrm{i}(\theta_{j}-\theta_{k})}/d$, i.e., $\rho$ is a pure state with equal diagonal elements; 3). $g$ is invariant under permutations of the $d$ path labels; 4). $g$ is convex.
Again, we use  relative entropy as the function $g$, based on the  measurements of $\{|i\rangle\langle i|\}_{i=1}^d$, we define the wave property as
\begin{eqnarray}\label{wav}
\mathcal{W}^{\tilde{\mathcal{B}}}(\rho):=C(\rho)-C[\Phi^{\tilde{\mathcal{B}}}(\rho)].
\end{eqnarray}
From Eq. (\ref{th1}), one obtains that it is just the coherence with respect to the incoherent basis $\tilde{\mathcal{B}}$, which depends on the measurements $\{|i\rangle\langle i|\}_{i=1}^d$ corresponding to multi-slit of the interference experiment.  Therefore, $\mathcal{W}^{\tilde{\mathcal{B}}}(\rho)$ meets all the items except for the third one. Since the permutation of the diagonal entries of density matrix $\rho$ does not alter the eigenvalues of $\rho^{\mathrm{diag}}$, the third item is proved.

Suppose $\vec{x}$ and $\vec{y}$ are vectors with their components in decreasing order, i.e., $x_1\geq x_2\geq\cdots\geq x_d$ and $y_1\geq y_2\geq\cdots\geq y_d$. If $\vec{x}$ and $\vec{y}$ satisfy $\sum_{i=1}^n x_i\leq \sum_{i=1}^ny_i$ for $i=1,2,\cdots,n$ with equality at $n=d$, then we say that $\vec{x}$ is majorized by $\vec{y}$ and denote $\vec{x}\prec\vec{y}$. Set $\vec{\rho}^\mathrm{diag}$ and $\vec{\frac{I}{d}}$ are vectors with the diagonal elements of state $\rho$ and $\frac{I}{d}$. Since $\vec{\rho}^\mathrm{diag}\prec \vec{\frac{I}{d}}$ and von Neumann entropy $S$ is Schur concave function, this means that $I/d$ is the incoherent state with larger wave property in the incoherent basis $\tilde{\mathcal{B}}$. Therefore, the longer the distance between $\Phi^{\tilde{\mathcal{B}}}(\rho)$ and $I/d$, the lesser the wave property.
As it was required, the maximum $\mathcal{W}^{\tilde{\mathcal{B}}}(\rho)=\log_2 d$ holds for pure states with equal diagonal elements under the  basis $\tilde{\mathcal{B}}$. 
If $\rho=I/d$, then one obtains $\mathcal{W}^{\tilde{\mathcal{B}}}(\rho)=P^{\tilde{\mathcal{B}}}(\rho)=0$.
On the other hand, if $\rho\ne I/d$, we can get the following trade-off relation
\begin{eqnarray}\label{tradeoff}
P'(\rho)+W'(\rho)=1,
\end{eqnarray}
where $P'(\rho)=P^{\tilde{\mathcal{B}}}(\rho)/(\log_2d-S(\rho))$ and $W'(\rho)=\mathcal{W}^{\tilde{\mathcal{B}}}(\rho)/(\log_2d-S(\rho))$. In the following, we give the interpretation of $S(\rho)$ in Eq. (\ref{tradeoff}) in wave-particle duality.

For any state $\rho$, one can always finds a reference system $R$ such that $|\Psi\rangle\in H\otimes H_R$, where $H_R$ is the Hilbert space and $\rho=\mathrm{Tr}_R|\Psi\rangle\langle\Psi|$. 
We define $E_f(|\Psi\rangle)=S(\rho)$ as the entanglement between system $H$ and $H_R$, for example, one can take $E_f(|\Psi\rangle)$ as entanglement of formation. For pure state $E_f(|\Psi\rangle)=S(\rho)$, while for mixed state, one can define $E_f(\rho)=\inf_{\{P_i, |\Psi_i\rangle\}}\sum P_iE_f(|\Psi_i\rangle)$, where the infinum is take over the ensembles $\{P_i,|\Psi_i\rangle\}$ such that $\sum_i P_i|\Psi_i\rangle=\rho_{HR}$ with $\rho=\mathrm{tr}_R\rho_{HR}$.
Then one can get the following conclusion
\begin{eqnarray}\label{main}
\mathcal{W}^{\tilde{\mathcal{B}}}(\rho)+P^{\tilde{\mathcal{B}}}(\rho)+E_f(|\Psi\rangle)=\log_2d,
\end{eqnarray}
where $\log_2d$ can be viewed as the maximum information available in the quantum system $H$. In Eq.(\ref{main}), we give an equality for wave-particle duality, and the right side of Eq.(\ref{main}) is content only depends on the dimension of the system. Thus, we have the following theorem.

{\bf Theorem 4}. Given a state $\rho$, for any basis $\mathcal{B}$, we have the following relation
\begin{eqnarray}\label{th4}
\mathcal{W}^{\mathcal{B}}(\rho)+P^{\mathcal{B}}(\rho)+E_f(|\Psi\rangle)=\log_2d,
\end{eqnarray}
where $d$ is the dimension of the system, $|\Psi\rangle$ is the purification of $\rho$. 

If $\rho_{AB}$ is a pure state, denote as $|\psi\rangle_{AB}$, from Eq. (\ref{th4}), one gets
\begin{eqnarray}\label{m1}
\mathcal{W}^{\tilde{\mathcal{B}}}(\rho_A)+P^{\tilde{\mathcal{B}}}(\rho_A)+E_f(|\psi\rangle_{AB})
=\log_2d_A,
\end{eqnarray}
where $d_A$ is the dimension of subsystem $A$. Similarly, one obtains
\begin{eqnarray}\label{m2}
\mathcal{W}^{\tilde{\mathcal{B}}}(\rho_B)+P^{\tilde{\mathcal{B}}}(\rho_B)+E_f(|\psi\rangle_{AB})
=\log_2d_B.
\end{eqnarray}
Combining with Eq. (\ref{m1}) and Eq. (\ref{m2}), we have 
\begin{eqnarray*}\label{m3}
\log_2d_A+\mathcal{W}^{\tilde{\mathcal{B}}}(\rho_B)+P^{\tilde{\mathcal{B}}}(\rho_B)
=\log_2d_B+\mathcal{W}^{\tilde{\mathcal{B}}}(\rho_A)+P^{\tilde{\mathcal{B}}}(\rho_A).
\end{eqnarray*}
If $d_A=d_B$, we have $\mathcal{W}^{\tilde{\mathcal{B}}}(\rho_A)+P^{\tilde{\mathcal{B}}}(\rho_A)=\mathcal{W}^{\tilde{\mathcal{B}}}(\rho_B)+P^{\tilde{\mathcal{B}}}(\rho_B)$. If $d_A< d_B$, we have $\mathcal{W}^{\tilde{\mathcal{B}}}(\rho_A)+P^{\tilde{\mathcal{B}}}(\rho_A)<\mathcal{W}^{\tilde{\mathcal{B}}}(\rho_B)+P^{\tilde{\mathcal{B}}}(\rho_B)$.

If $\rho_{AB}$ is a mixed state, one can always finds a reference system $C$ such that $\rho_{AB}=\mathrm{Tr}_C|\psi\rangle_{ABC}\langle\psi|$. Suppose $d_C\geq d_Ad_B\geq d_B\geq d_A$ 
Then using Eq. (\ref{main}) again, one obtains the following equations,
\begin{eqnarray}\label{m4}
\mathcal{W}^{\tilde{\mathcal{B}}}(\rho_A)+P^{\tilde{\mathcal{B}}}(\rho_A)+E_f(|\psi\rangle_{A|BC})
=\log_2d_A,
\end{eqnarray}
\begin{eqnarray}\label{m5}
\mathcal{W}^{\tilde{\mathcal{B}}}(\rho_B)+P^{\tilde{\mathcal{B}}}(\rho_B)+E_f(|\psi\rangle_{B|AC})
=\log_2d_B,
\end{eqnarray}
\begin{eqnarray}\label{m6}
\mathcal{W}^{\tilde{\mathcal{B}}}(\rho_C)+P^{\tilde{\mathcal{B}}}(\rho_C)+E_f(|\psi\rangle_{C|AB})
=\log_2d_C,
\end{eqnarray}
\begin{eqnarray}\label{m7}
\mathcal{W}^{\tilde{\mathcal{B}}}(\rho_{AB})+P^{\tilde{\mathcal{B}}}(\rho_{AB})+E_f(|\psi\rangle_{AB|C})
=\log_2d_{A}d_B,
\end{eqnarray}
Combining with Eq.(\ref{m4}), Eq.(\ref{m5}) and Eq.(\ref{m7}), one gets  
\begin{eqnarray*}\label{}
\mathcal{W}^{\tilde{\mathcal{B}}}(\rho_A)+P^{\tilde{\mathcal{B}}}(\rho_A)&+&\mathcal{W}^{\tilde{\mathcal{B}}}(\rho_B)+P^{\tilde{\mathcal{B}}}(\rho_B)\nonumber\\
&\leq& \mathcal{W}^{\tilde{\mathcal{B}}}(\rho_C)+P^{\tilde{\mathcal{B}}}(\rho_C),
\end{eqnarray*}
since $E_f(|\psi\rangle_{A|BC})+E_f(|\psi\rangle_{B|AC})\geq E_f(|\psi\rangle_{C|AB})$, and the equation satisfied for $d_C=d_Ad_B$. One can see that the uncertainty of the wave and particle property is invariant under unitary operation and is bigger with the increasing of the dimension of system.
Combining with Eq.(\ref{m6}) and Eq.(\ref{m7}), we have 
\begin{eqnarray*}\label{}
\mathcal{W}^{\tilde{\mathcal{B}}}(\rho_{AB})+P^{\tilde{\mathcal{B}}}(\rho_{AB})\leq \mathcal{W}^{\tilde{\mathcal{B}}}(\rho_C)+P^{\tilde{\mathcal{B}}}(\rho_C),
\end{eqnarray*}
where the equation satisfied for $d_C=d_Ad_B$.

 For a single qubit can be written as $\rho=\frac{I+\vec{r}\cdot \vec{\sigma}}{2}$, where $\vec{r}=(x,y,z)$ is a real three-dimensional vector. Let $x=\frac{1}{\sqrt{2}},y=\frac{1}{\sqrt{3}},z=\frac{1}{\sqrt{6}}$, then, go through a bit flipping channel, i.e., $E_0=\sqrt{p}I,~E_1=\sqrt{1-p}X$, $X$ represents quantum NOT gate. One gets \begin{equation}
 \rho'= \left( \begin{array}{ccc} 
  \frac{1}{2}(1+\frac{2p-1}{\sqrt{6}}) & \frac{1}{2\sqrt{2}}+\frac{1-2p}{2\sqrt{3}} \mathrm{i} \\ 
   \frac{1}{2\sqrt{2}}-\frac{1-2p}{2\sqrt{3}} \mathrm{i}  &   \frac{1}{2}(1-\frac{2p-1}{\sqrt{6}})  \\
   \end{array} \right),
 \end{equation}
 where $\mathrm{i}=\sqrt{-1}$.
 The relations among wave property, particle property and entanglement with the reference system for initial state $\rho$ under bit flipping channel is shown in Fig. 5. One can see that the wave property  is more affected by the channel than the partical property, and wave property, particle property decrease with increasing entanglement.
\begin{figure}
	\centering
	\includegraphics[width=8cm]{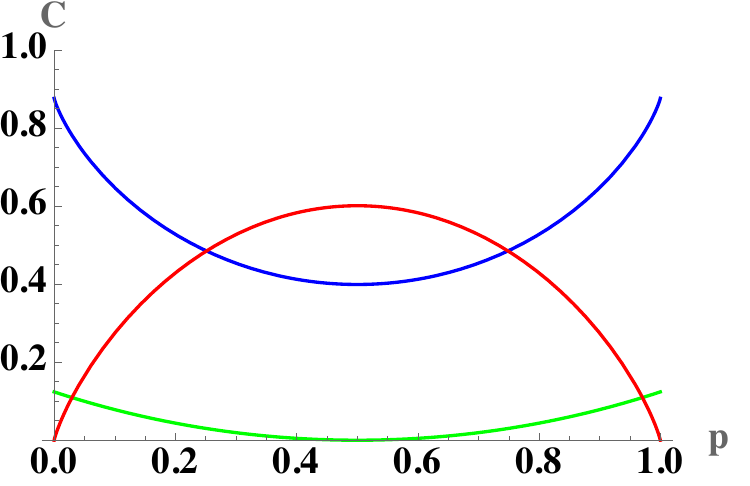}\\
	\caption{The blue solid line is wave property of $\rho'$, green solid line is particle property of $\rho'$ and red solid line is the entanglement with the reference system.}\label{2}
\end{figure}

\section{Hierarchical relationship of quantum correlation}
Similar to Eq. (\ref{rela}), the distance-based quantifiers for entanglement and discord can be defined as
\begin{eqnarray}\label{ent}
E(\rho)=\min_{\delta\in \mathcal{S}} S(\rho || \delta),\\
D(\rho)=\min_{\gamma\in \mathcal{Z}} S(\rho || \gamma),
\end{eqnarray}
where $\mathcal{S}$ and $\mathcal{Z}$ denote the sets of separable and zero-discord states, respectively.
Quantum discord can either be defined with respect to a particular subsystem, or symmetrically concerning all subsystems. Here, we consider the symmetrical one.
Since the original incoherent states $\mathcal{I}$ are diagonal states defined in a predetermined orthogonal basis, and zero-discord states are a subset of separable states, the inclusion of sets clearly appears: $I/d\in \mathcal{I}\subset \mathcal{Z} \subset \mathcal{S} $.
Then we have the following natural ordering of quantum correlation: 
\begin{eqnarray}\label{qcc}
C(\rho)\geq C^\mathcal{B}(\rho)\geq D(\rho)\geq E(\rho),
\end{eqnarray}
which signifies that, despite that almost all quantum states
exhibit nonzero discord \cite{fa}, BI quantum coherence is a more ubiquitous manifestation of quantum correlations. 
Now we use an example to illustrate (\ref{qcc}) more clearly. Consider the Bell-diagonal states,
$\rho_{AB}=\frac{1}{4}(I\otimes I+\sum_{j=1}^3c_j\sigma_j\otimes\sigma_j)=\sum_{ab}\lambda_{ab}|\beta_{ab}\rangle\langle\beta_{ab}|$, where $\sigma_j$ are the standard Pauli matrices.
One gets $C(\rho_{AB}) =2-H(\lambda_{ab})$, $C^{\tilde{\mathcal{B}}}(\rho_{AB}) =2-H(\lambda_{ab})-\sum_{j=1}^2\frac{(1+(-1)^jc_3)}{2}\log_2(1+(-1)^jc_3)$ and $D(\rho_{AB})=2-H(\lambda_{ab})-\sum_{j=1}^2\frac{(1+(-1)^jc)}{2}\log_2(1+(-1)^jc)$, where $c=\max\{|c_1|,|c_2|,|c_3|\}$, see Appendix C for detailed derivations.
From the results in \cite{vv}, one has $E(\rho_{AB})=1+\lambda_{00}\log_2\lambda_{00}+(1-\lambda_{00})\log_2(1-\lambda_{00})$ for $\lambda_{00}\geq \frac{1}{2}$, and $E(\rho_{AB})=0$, when $\lambda_{00}\in [0,\frac{1}{2}]$.

In \cite{yy}, the authors proposed a basis-free measure of coherence by the minimization over all local unitary transformations $C^\mathrm{free}(\rho)=\min_{\vec{U}}C^\mathcal{B}(\vec{U}\rho\vec{U}^\dagger)$ with $\vec{U}=U_1\otimes U_2\otimes\cdots \otimes U_N$ possessing a local product structure.
Our definition of BI coherence is different from the one in \cite{yy}. In fact, $C(\rho)$ in Eq. (\ref{bn}) is equivalent to
$C_U(\rho)=\max_UC^\mathcal{B}(U\rho U^\dagger)$ with the maximization over all global unitary transformations $U$, see proof in Appendix D. Thus one has $C(\rho)=C_U(\rho)\geq C^\mathcal{B}(\rho)\geq C^\mathrm{free}(\rho)$.

Finally, BI coherence can be used to estimate the transformation rate $R(\rho\to |\psi\rangle)$.  In Ref. \cite{mh4}, it turns out that the information $n(\log_2 d-S(\rho))$ (i.e., $nC(\rho)$ defined in Eq. (\ref{bn})) is just the number of pure states which can be distilled from $n$ copies of $\rho$. One can also show that the converse protocol is possible: i.e., to create $n$ copies of state $\rho$ one needs approximately $nC(\rho)$ pure qubits and $nS(\rho)$ qubits of maximally mixed state.
BI coherence is also related to the average work extracted from the qubit in quantum thermodynamics. In \cite{lm}, the authors present that the amount of work extractable from a given state $\rho$ in contact with a thermal reservoir at a fixed temperature is $W=k_BT(\log_2d-S(\rho))=k_BTC(\rho)$, where $k_B$ and $T$ are the Boltzmann constant and temperature of the system. Using the projection onto the energy eigenstates $\{\Pi_k^H\}$ of the system Hamiltonian $H=\sum_k E_k\Pi_k^H$, one obtians $\rho\to \sigma^H=\sum_k\mathrm{Tr}(\rho\Pi_k^H )\Pi_k^H$.  It has been shown in \cite{kp} that the average work extracted from the qubit systems during the conversion of quantum coherence into work is $W=k_BT(S(\sigma^H)-S(\rho))$. Submitting Eq. (\ref{th1}) into the above equality, the work extracted from the qubit systems can be written as $W=k_BT(C(\rho)-C(\sigma^H))$.
Hence we find that the average work extracted from the qubit is just the difference between the BI coherence in the state $\rho$ and its projected state $\sigma^H$.

\section{Conclusion} 
We have analyzed quantum coherence in a basis-independent manner based on the relative entropy, which is in contrast to the original coherence studied as a property regarding a particular chosen basis. 
We have shown that the incoherent state should be invariant under arbitrary unitary transformations, that is the maximally mixed state. This gives a simpler way to quantify coherence since a minimization over free states is not necessary.
Then we have given the relationship between the basis-independent and the basis-dependent approaches. By using the eigenvectors of the observable as the measurement, we have gotten the difference of BI coherence between the original state and post-measurement state is just the original coherence for the observer's choice (particular basis). 
As an application, we have shown that the wave-particle uncertainty relation being related to the entanglement between system and reference system, and the sum of them is the maximum information available in a quantum system, which only depends on the dimension of system. The uncertainty of the wave-particle property is based on the dimension of system, and the uncertainty relation increase with the increase of the system dimension.  For a state $\rho$, the best case to obtain the entirely BI coherence of $\rho$ is to use two sets of measurements.
 We also have explored the hierarchical relationship of the basis-independent quantum coherence along with the various quantum correlations.

\bigskip
\noindent{\bf Acknowledgments}\, \,
This work was supported in part by the National Natural Science Foundation of China (NSFC) under Grants 12301582; Guangdong Basic and Applied Basic Research Foundation under Grants No. 2024A1515030023; Start-up Funding of Dongguan University of Technology No. 221110084; Start-up Funding of Guangdong Polytechnic Normal University  No. 2021SDKYA178.


\begin{thebibliography}{99}
\bibitem{cr} C. Radhakrishnan, M. Parthasarathy, S. Jambulingam, and T. Byrnes, Distribution of Quantum Coherence in Multipartite Systems, Phys. Rev. Lett. 116, 150504 (2016).
\bibitem{aes} A. Streltsov, E. Chitambar, S. Rana, M. Bera, A. Winter, and M. Lewenstein, Entanglement and coherence in quantum state merging, Phys. Rev. Lett. 116, 240405 (2016).
\bibitem{zfl}F. L. Zhang and T. Wang, Intrinsic coherence in assisted sub-state discrimination, Europhysics Letters 117, 10013 (2017).
\bibitem{gbf} G. Karpat, B. Cakmak, and F. Fanchini, Quantum coherence, and uncertainty in the anisotropic XY chain, Phys. Rev. B 90, 104431 (2014).
\bibitem{agb} A. Malvezzi, G. Karpat, B. Cakmak, F. Fanchini, T. Debarba, and R. Vianna, Quantum correlations and coherence in spin-1 Heisenberg chains, Phys. Rev. B 93, 184428 (2016).
\bibitem{cx} J. Q. Cheng and J. B. Xu, Multipartite entanglement, quantum coherence, and quantum criticality in triangular and Sierpiński fractal lattices, Phys. Rev. E 97, 062134 (2018)
\bibitem{szh} Y. T. Sha, Y. Wang, Z. H. Sun, and X. W. Hou, Thermal quantum coherence and correlation in the extended XY spin chain, Annals of Physics 392 (2018).

\bibitem{pmb}T. Baumgratz, M. Cramer, and M. B. Plenio, Quantifying Coherence, Phys. Rev. Lett. 113, 140401 (2014).
\bibitem{cb} C. Radhakrishnan, M. Parthasarathy, S. Jambulingam and T. Byrnes, Distribution of quantum coherence in multipartite systems, Phys. Rev. Lett., 116, 150504 (2016).
\bibitem{gc} E. Chitambar and G. Gour, Comparison of incoherent operations and measures of coherence, Phys. Rev. A, 94 052336 (2016).
\bibitem{jzx1} Jin, ZX., Yang, LM., Fei, SM. et al. Maximum relative entropy of coherence for quantum channels. Sci. China Phys. Mech. Astron. 64, 280311 (2021). 
\bibitem{rae}  A. E. Rastegin, Quantum-coherence quantifiers based on the Tsallis relative $\alpha$ entropies, Phys. Rev. A, 93, 032136 (2016).
\bibitem{rpp} S. Rana, P. Parashar, and M. Lewenstein, Trace-distance measure of coherence, Phys. Rev. A, 93, 012110 (2016).
\bibitem{ssd} A. Streltsov, U. Singh, H. S. Dhar, M. N. Bera, and G. Adesso, Measuring quantum coherence with entanglement, Phys. Rev. Lett., 115, 020403 (2015).
\bibitem{reb} C. Radhakrishnan, I. Ermakov and T. Byrnes, Quantum coherence of planar spin models with Dzyaloshinsky-Moriya interaction, Phys. Rev. A, 96, 012341 (2017).
\bibitem{jzx} Z. X. Jin, S. M. Fei, Quantifying quantum coherence and nonclassical correlation based on Hellinger distance, Phys. Rev. A, 97, 062342 (2018).
\bibitem{huml}Hu, M. L. et al. Quantum coherence and geometric quantum discord. Phys. Rep. 762-764, 1–100 (2018).
\bibitem{nbc} C. Napoli, T. R. Bromley, M. Cianciaruso, M. Piani, N. Johnston and G. Adesso, Robustness of coherence: an operational and observable measure of quantum coherence, Phys. Rev.
Lett., 116, 150502 (2016).
\bibitem{wyd} A. Winter and D. Yang, Operational resource theory of coherence, Phys. Rev. Lett., 116, 120404 (2016).
\bibitem{mx}X. Yuan, H. Zhou, Z. Cao, and X. Ma, Intrinsic randomness as a measure of quantum coherence, Phys. Rev. A, 92, 022124 (2015).
\bibitem{csr} E. Chitambar, A. Streltsov, S. Rana, M. Bera, G. Adesso, and M. Lewenstein, Assisted distillation of quantum coherence, Phys. Rev. Lett., 116, 070402 (2016).

\bibitem{gan} I. Georgescu, S. Ashhab and F. Nori, Quantum simulation, Rev. Mod. Phys., 86, 153 (2014).
\bibitem{yy}Y. Yao, X. Xiao, L. Ge, and C. P. Sun, Quantum coherence in multipartite systems, Phys. Rev. A 92, 022112 (2015).
\bibitem{glm} V. Giovannetti, S. Lloyd, and L. Maccone, Quantum-enhanced measurements: beating the standard quantum limit, Science, 306, 1330 (2004).



\bibitem{dm} R. Demkowicz-Dobrzanski and L. Maccone, Using entanglement against noise in quantum metrology, Phys. Rev. Lett., 113, 250801 (2014).
\bibitem{eak} A. K. Ekert, Quantum cryptography based on Bell's theorem, Phys. Rev. Lett., 67, 661 (1991).
\bibitem{ycs} C. S. Yu, Quantum coherence via skew information and its polygamy, Phys. Rev. A 95, 042337 (2017).

\bibitem{yy}Y. Yao, X. Xiao, L. Ge, and C. P. Sun, Quantum coherence in multipartite systems, Phys. Rev. A 92, 022112 (2015).



\bibitem{lsl1} S. Luo and Y. Sun, Quantum coherence versus quantum uncertainty, Phys. Rev. A 96, 022130 (2017).



\bibitem{yang} C. S. Yu, S. R. Yang, and B. Q. Guo, Total quantum coherence and its applications, Quantum Information Processing 15, 3773 (2016).
\bibitem{scp}Y. Yao, G. Dong, X. Xiao, and C. Sun, Frobenius-norm-based measures of quantum coherence and asymmetry, Scientific reports 6, 32010 (2016).
\bibitem{hml}M. L. Hu, S. Q. Shen, and H. Fan, Maximum coherence in the optimal basis, Phys. Rev. A 96, 052309 (2017).
\bibitem{ma} Z. H Ma, J. Cui, Z. Cao, S. M. Fei, V. Vedral, T. Byrnes and C. Radhakrishnan, Operational advantage of basis-independent quantum coherence, EPL, 125, 50005 (2019)
\bibitem{durr} S. Durr, Quantitative wave-particle duality in multibeam interferometers, Phys. Rev. A 64, 042113 (2011).
\bibitem{sm} S. Camalet, Internal entanglement and external correlations of any form limit each other, Phys. Rev. Lett. 121, 060504 (2018).
\bibitem{la} A. Luis and L. Monroy, Nonclassicality of coherent states: Entanglement of joint statistics, Phys. Rev. A 96, 063802 (2017).

\bibitem{fa} A. Ferraro, L. Aolita, D. Cavalcanti, F. M. Cucchietti, and A. Acin, Almost all quantum states have nonclassical correlations, Phys. Rev. A 81, 052318 (2010).
\bibitem{mh4}M. Horodecki, P. Horodecki and J. Oppenheim, Reversible transformations from pure to mixed states and the unique measure of information, Phys. Rev. A 67, 062104 (2003).
\bibitem{lm}M. Lostaglio, D. Jennings and T. Rudolph, Description of quantum coherence in thermodynamic processes require constraints beyond free energy, Nat. Commun., 6, 6383 (2015).



\bibitem{kp} P. Kammerlander and J. Anders, Coherence and measurement in quantum thermodynamics, Sci. Rep., 6, 22174 (2016).








\bibitem{lsl} S. Luo, Quantum discord for two-qubit systems, Phys. Rev. A 77, 042303 (2008).

\bibitem{vv}  V. Vedral, M.B. Plenio, M.A. Rippin and P.L. Knight, Phys. Rev. Lett. 78, 2275 (1997).














\end{thebibliography}

\begin{thebibliography}{99}



\end{thebibliography}
\end{document}